# A High-confidence Cyber-Physical Alarm System: Design and Implementation

Longhua Ma[1,2], Tengkai Yuan[1], Feng Xia[3], Ming Xu[1], Jun Yao[1], Meng Shao[4]
[1]Department of Control Science and Engineering, Zhejiang University, Hangzhou 310027, China
[2]School of Aeronautics and Astronautics, Zhejiang University, Hangzhou 310027, China
e-mail: lhma@iipc.zju.edu.cn
[3]School of Software, Dalian University of Technology, Dalian 116620, China
e-mail: f.xia@ieee.org
[4]Computer Centre, Hangzhou First People's Hospital, Hangzhou 310006, China

*Abstract*—Most traditional alarm systems cannot address security threats in a satisfactory manner. To alleviate this problem, we developed a high-confidence cyber-physical alarm system (CPAS), a new kind of alarm systems. This system establishes the connection of the Internet (i.e. TCP/IP) through GPRS/CDMA/3G. It achieves mutual communication control among terminal equipments, human machine interfaces and users by using the existing mobile communication network. The CPAS will enable the transformation in alarm mode from traditional one-way alarm to two-way alarm. The system has been successfully applied in practice. The results show that the CPAS could avoid false alarms and satisfy residents' security needs.

## I. INTRODUCTION

With the development of society, social security has become increasingly demanding, and the traditional Passive-Alarm (PA) mode has been far from meeting our security need. As we know, PA mode will not trigger an alert signal until the alarm condition occurs, which can not alleviate users' security threats. Meanwhile, the Cyber-Physical System (CPS) is becoming an increasingly hot topic, and even being as a national development strategy. For example, the US National Science Foundation (NSF) has identified CPS a key area of research [1]. There are many fields involved in CPS, such as Smart Home, Smart Medical, Intelligent Transportation, Intelligent Power Grid, etc [2,3]. In this paper, a high confidence Cyber-Physical Alarm System (CPAS) is presented, which will achieve smart alarm, and bring unprecedented challenges and opportunities for the security industry simultaneously.

CPS is a system featuring a tight combination of, and coordination between, the system's computational and physical elements [4], and it is recognized as the third wave of the world information industry followed by the Computer, Internet and Mobile Communication Network. Unlike more traditional embedded systems, a fill-fledged CPS is typically designed as a network of interacting elements instead of as standalone devices [5]. While CPAS is a system in which the terminal equipment (TE), intelligent and integrated, is able to connect wireless sensors, transmit alert information and images via the Internet. What's more, the monitoring alarm service based on network platform is included. There is no doubt that the traditional one-way alarm system will be replaced by CPAS, and the social security protection will get into a new stage.

The rest of the paper is organized as follows. Section II presents the model of CPAS. The requirements and features of the system are specified. In Section III, we discuss the major challenges that need to be addressed when realizing a CPAS. Section IV illustrates how we design and implement the system. An application case study is presented in Section V. We finally conclude the paper in Section VI.

## II. CPAS MODELING

### A. System model

In our daily life, the telephone communication network serves for communications among people. Sensor network is the perceptible network, serving for communications among objects. And the Internet, a virtual information space, is the information-sharing network. The three are information-transmission network, information-perception network and information-sharing network respectively, while CPS is the integration of the three. The CPS bridges the virtual world and the real word much closer with the combination of sensors, intelligent terminals and the Internet. Through embedding a variety of sensors, such as images, humidity, temperature, infrared, etc., into different objects or the environment, the information of objects or environment will be combined with the Internet by intelligent network terminal, and then reaches to the integration of human society and physical system.

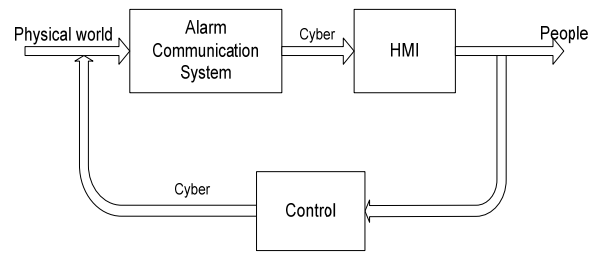

Figure 1. Simplified CPAS model

In our approach, the alarm TE, integrated with infrared (IR) and video surveillance, will trigger video surveillance as soon as the IR probe detects an alert, and transmit the site image in real time. Then, an alarm task is finished.

Figure 1 illustrates a simplified CPAS model. The Alarm Communication System (ACS) is a TE, used to detect the physical world. Through the Human Machine Interface (HMI), we may observe information about the physical world, which is transmitted via the ACS. We could also control and manage the physical world in turn through the network. In this way, the mutual dialogue among the physical world, cyber world and human world is achieved, in which a high confidence CPAS will be formed.

*B. Requirements*

To establish an effective CPS, the following two factors need to be considered.

(1) Scale. It is only when the scale of smart TEs reaches to a certain size that the intelligent aspect of an object can take effect. For example, if there are one million cars, and only 10 thousands of them are embedded with intelligent systems, then it will be very difficult, if not impossible, to form an intelligent transportation system.

(2) Mobility. Objects are usually not static, but in a state of movement. On the other hand, the communication control among objects is supposed to be achieved at any time, whether in motion or even under high speed. As a consequence, a mobile CPS is often required.

Nowadays, the wireless mobile communication network (WMCN) has covered most areas in China, from bustling cities to remote rural areas, from island to Mount Everest. Considering this fact, our embedded ACS is designed to access to Internet by the way of GPRS/CDMA. Compared to 3G or 4G, GPRS(2.5G)/CDMA(2.75G) has wider coverage of the WMCN, which is more accepted by users. Of course, we will try to use 3G/4G when the coverage of 3G/4G is more widespread in future.

*C. Features*

Unlike traditional alarm systems, the CPAS has its own functions. Intelligent alarm host may connect to a variety of wireless sensors to achieve the perception of external environment. For example, in case of fire, the temperature and smoke sensors will transmit related data immediately to the alarm host in a wireless way at the beginning stage of fire. After receiving the data, the alarm host will make a determination whether to send out a warning alert, so that the user may take timely actions to reduce loss as much as possible.

By using low-power microprocessor MSPF430/ARM7, the CPAS triggers video surveillance through IR probe and transmits monitoring images in real time, which facilitates green monitoring alarm. In brief, the new generation alarm system (i.e. CPAS) has these characteristics: small-size, low-cost, low power dissipation, and combination of monitoring, image analysis, intelligent processing, siren, active-alarm (AA) and other intelligent functions.

III. CHALLENGES

One of the challenges to make a CPAS to work is to urge the embedded alarm system to access to the Internet. It is necessary to examine both advantages and disadvantages of different ways of entering the Internet. For example, we should take real-time, cost, reliability, etc., into consideration when transmitting alert information.

Another challenge is that HMI must support multi-threads, more specifically, thousands and even ten thousands of threads at run time. This is because the HMI usually needs to receive alert information from a large number of alarm TEs placed in various environments, where a Distributed Cyber- Physical System (DCPS) [6] will be constructed. In addition, the HMI is supposed to be in connection with users, in order to ensure that the corresponding user will be informed when an alarm occurs. Thus a Distributed Cyber-Physical Alarm System (DCPAS) may be formed in a certain place. Figure 2 shows some kind of DCPAS.

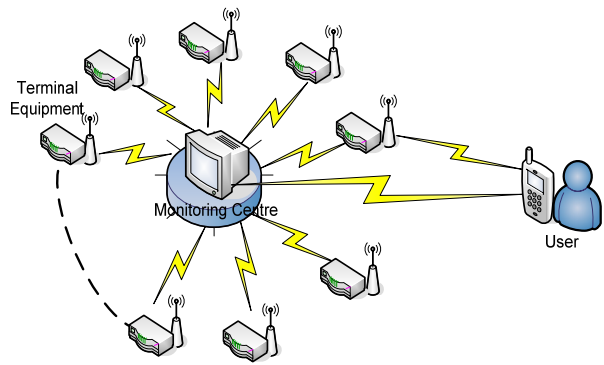

Figure 2. DCPAS schematic diagram

A third challenge is the communication protocol among the TE, HMI and user. As in other networked systems, the protocol has a direct impact on the reliability of the CPAS system. A widely-accepted, unified protocol is the premise of implementing a large-scale CPAS.

IV. SYSTEM DESIGN AND IMPLEMENTATION

*A. TCP/IP Connection of TEs*

A wireless communication module, embedded with TCP/IP protocol, is required for the TE to establish TCP/IP connection. Examples of such modules include GPRS wireless module MC55 by Siemens, 3G-WCDMA wireless module MU103 by Huawei, etc.

The transmission rate of GPRS already satisfies the requirements of civil embedded security systems, whose average rate ranges from 20kb/s to 30kb/s, and its maximum theoretical speed may reach to 171.2kb/s. Although the rate of CDMA is much higher than that of GPRS, the expense of CDMA is much higher too, and the expense of 3G is far higher. Therefore, GPRS is much more popular before the expense of CDMA or 3G cuts down. The embedded alarm system selects GPRS to establish the TCP/IP connection, which yields lower cost and makes it more affordable for users. The approach for the alarm TE to establish a TCP/IP connection to access the Internet is through AT commands control. Taking MC55 for example, Figure 3 illustrates its process of TCP/IP connection.

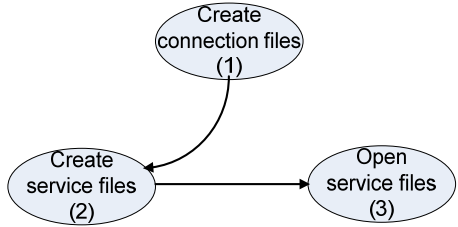

Figure 3. Process of TCP/IP connection

To ensure the reliability of connected TCP/IP, much experience and technology are required. There are two factors closely related to reliability: (1) hardware: the timing condition of the module itself; and (2) software: the microcontroller is supposed to activate the wireless module continuingly through AT commands so that the module will not disconnect TCP/IP abnormally. Taking again MC55 for instance, Figure 4 illustrates the timing of power-on process of the GPRS wireless module [7].

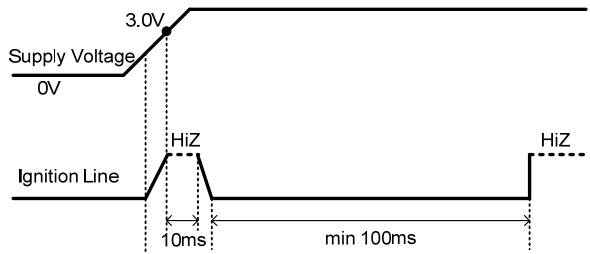

Figure 4. Timing of power-on process of MC55

The ignition line should not be switched low until 10ms delay after the wireless module is powered up. The ignition line should be maintained in the state of low-pulled more than 100ms, so that the module will start to work effectively. In addition, we also need to switch the ignition line low and high, i.e. equal to HiZ (high impedance), now and then during run time. By doing this, the module will continue to work spontaneously as soon as the module fails. From the software perspective, the wireless module would disconnect TCP/IP abnormally if we do not transmit data through GPRS for a long time. Thus the microcontroller needs to control wireless module to enable GPRS at periodic intervals to ensure that the connection of TCP/IP is normal. For instance, we may send a packet through GPRS every minute.

### B. Multi-threading Solution

Thread is the basic atomic unit of execution of a procedure, and a process can be composed of multiple threads. The implementation strategy is to divide a process into multiple threads, and then let them execute concurrently and asynchronously in order to improve operational efficiency. Concurrent execution does not mean that all threads run at the same time (occupying CPU simultaneously), but only one thread is admitted to occupy CPU at any time. Since some threads compete for CPU more frequently, they seem to run simultaneously. In distributed programming, the proper use of threads can be very good to enhance the performance and efficiency of application procedure.

However, multithreading gives rise to new problems. For example, if many threads must occupy a certain resource or several resources, this may possibly cause threads to be mixed, or even collapsed. To avoid these problems, the Resource Allocation Algorithm (RAA) or Time-Slice Rotation Scheduling Algorithm (TSRSA) [8] is needed. In this work, we adopt TSRSA. The basic idea behind this algorithm is: firstly, divide the processing time of CPU into a slice of time; then, each thread in ready queue takes turns to use CPU resources according to assigned time slice. When the allotted time slice runs out, the thread will be forced to give up CPU, and re-enter the end line of the ready queue to wait for the next scheduling. Meanwhile, the process of scheduling goes to select the first thread in the ready queue, and allocates time slice to it. The following is a simplified mathematical model of TSRSA, taking $n$ threads for instance:

$$P = \sum_{i=1}^{n} x_i T_i \quad (1)$$

$$s.t. \begin{cases} \sum_i x_i = 1 \\ x_i \geq x_{i+1} \end{cases} \quad (2)$$

where $P$ stands for process, $T_i$ is the $i$-th thread. The $P$ is dynamically changing, since the first thread will go out of the schedule of $\sum_{i=1}^{n} x_i T_i$ as soon as it is finished. For this reason, the conditions given in (2) can ensure that $P$ will always deal with the first thread of the current ready queue.

### C. Communication Control

The most significant difference between CPAS and traditional alarm systems is reflected in the communication control. We achieve the communication control between TE and HMI based on GPRS, communication between TE and user based on SMS. What's more, users can control the TE and check the current status of TE in turn according to the agreed protocol through SMS, and security guards could also use HMI to control the TE and query the current state of TE in accordance with agreed protocol through GPRS. It can be seen that the CPAS has obvious advantages, and its two-way communication control can relieve the users' security threats completely.

Considering the protection of the CPS, much effort has been done in reliability (the protection against unpredictable failures) [9,10,11]. Our communication control among the TE, HMI and user may face information transmission security problems. Therefore, the protection against malicious cyber attacks should be concerned. Secure control had been described in e.g. [12], which is suitable for our CPAS to protect the communication control.

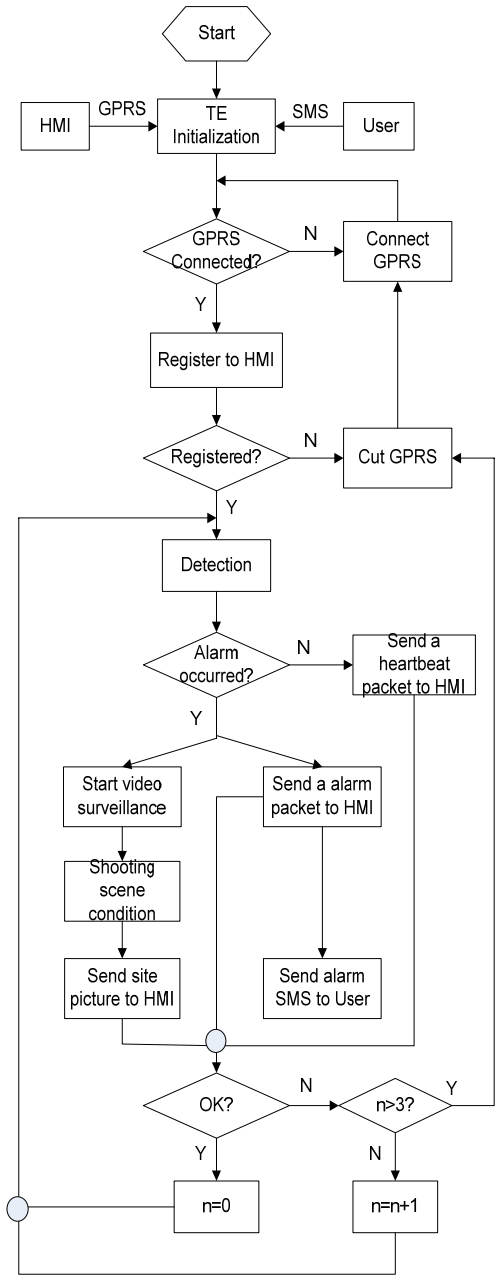

Figure 5. Process of communication control

Figure 5 shows the process of communication control among the TE, HMI and user, where *n* stands for the number of GPRS sending failures. To alleviate the problem of communication failures caused by network congestion, the intelligent TE will disconnect GPRS independently when *n*>3, and then reconnect GPRS immediately to ensure that the TE remains online almost all the time.

## V. AN APPLICATION CASE

The CPAS we developed have been applied in Dongyang City, China. The HMI succeeds to hold more than 2000 intelligent TEs from different users. Figure 6 gives the networked alarm interface, which comes from a testing experiment, with 45 TEs.

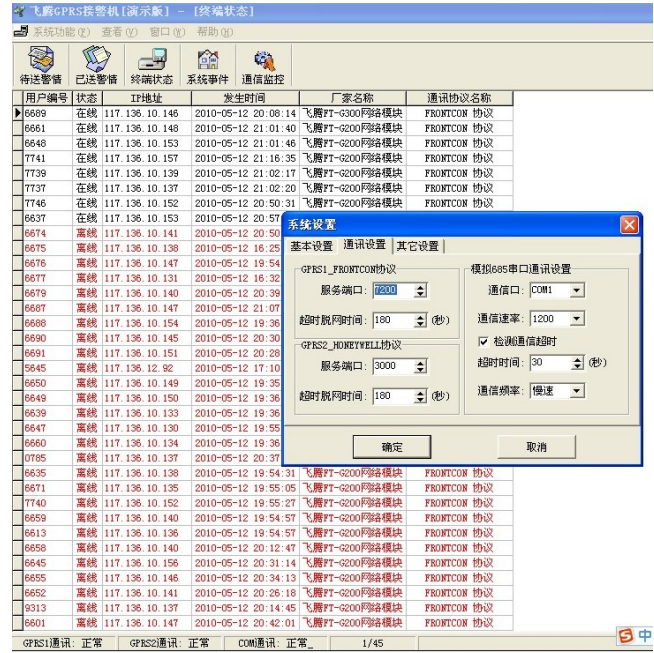

Figure 6. An interface of CPAS

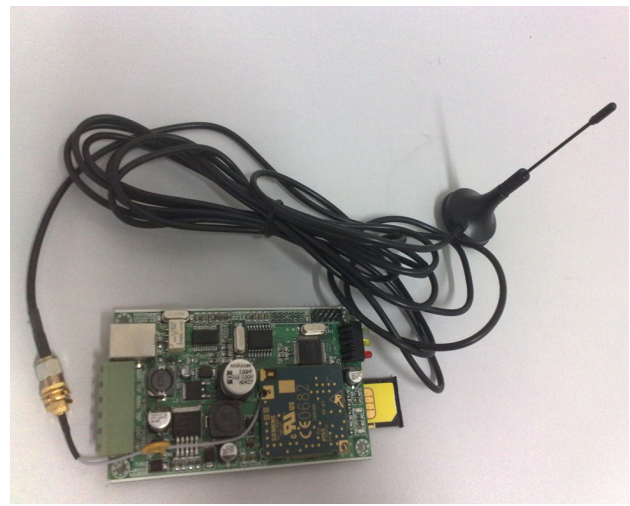

Figure 7. Intelligent alarm TE

As we can see from Figure 6, a state of "on-line" (indicated by black) means that the alarm TE is monitoring, while a state of "off-line" (in red) implies that the alarm TE is not working. If an alert occurs, the alarm will be finished in about 2 seconds. The HMI will prompt to the alarm interface, producing a sound to draw the security-guard's attention to take timely actions. At the same time, the intelligent TE will send an alert message to the user, thus achieving double protection. Figure 7 shows the intelligent alarm TE.

At present, the alarm TE can satisfy the users' needs without video surveillance. It is intuitive that more cost would be involved when starting video surveillance. The TEs have been on running for more than half a year, reliably and stably, in Dongyang City, where a small CPAS can succeed to operate and solve the problem of security threats for local residents.

## VI. Conclusions

The work presented in this paper represents a promising step towards the next generation of CPAS. We first modeled a simplified CPAS, and analyzed its requirements and key characteristics that are different from traditional alarm systems. We then discussed three main implementation challenges, and succeeded to solve them finally. A small CPAS was deployed in Dongyang City, China, demonstrating quite satisfactory performance. Empirical results show that the system we developed is well suited for satisfying civil security needs.

## Acknowledgment

This work is supported in part by Natural Science Foundation of China under Grant No. 60903153, Zhejiang Provincial Natural Science Foundation of China under Grant No. R1090052 and Grant No. Y108685, and the Fundamental Research Funds for the Central Universities.